\documentclass[twocolumn,superscriptaddress,amsmath,amsfonts,amssymb,preprintnumbers]{revtex4}
\pdfoutput=1
\usepackage[T1]{fontenc}\usepackage[latin1]{inputenc}
\usepackage{dcolumn,graphicx,color,booktabs,microtype}
\usepackage[charter]{mathdesign}
\renewcommand{\figurename}{Fig.}
\renewcommand{\tablename}{Table}
\makeatletter\renewcommand{\fnum@figure}[1]{\figurename~\thefigure.}\makeatother
\makeatletter\renewcommand{\fnum@table}[1]{\tablename~\thetable.}\makeatother
\usepackage[colorlinks,plainpages=false,linkcolor=black,urlcolor=blue,citecolor=black,pdfpagemode=UseNone,pdfstartview=FitBH]{hyperref}

\begin{document}
\title{$(\pi, \pi)$-electronic order in iron arsenide superconductors.}

\author{V.~B.~Zabolotnyy}
\affiliation{Institute for Solid State Research, IFW-Dresden,
P.O.Box 270116, D-01171 Dresden, Germany}
\author{D.~S.~Inosov}
\affiliation{Institute for Solid State Research, IFW-Dresden, P.O.Box 270116, D-01171 Dresden,
Germany} \affiliation{Max-Planck-Institute for Solid State Research, Heisenbergstraße 1, D-70569
Stuttgart, Germany}
\author{D.~V.~Evtushinsky}
\author{A.~Koitzsch}
\affiliation{Institute for Solid State Research, IFW-Dresden, P.O.Box 270116, D-01171 Dresden,
Germany}
\author{A.~A.~Kordyuk}
\affiliation{Institute for Solid State Research, IFW-Dresden, P.O.Box 270116, D-01171 Dresden, Germany}
\affiliation{Institute of Metal Physics of National Academy of Sciences of Ukraine, 03142 Kyiv, Ukraine}
\author{G.~L.~Sun}
\author{J.~T. Park}
\author{D.~Haug}
\author{V.~Hinkov}
\author{A.~V.~Boris}
\author{C.~T.~Lin}
\affiliation{Max-Planck-Institute for Solid State Research, Heisenbergstraße 1, D-70569 Stuttgart,
Germany}
\author{M.~Knupfer}
\affiliation{Institute for Solid State Research, IFW-Dresden, P.O.Box 270116, D-01171 Dresden,
Germany}
\author{A.~N.~Yaresko}
\affiliation{Max-Planck-Institute for Solid State Research, Heisenbergstraße 1, D-70569 Stuttgart,
Germany}
\author{B.~B\"{u}chner}
\affiliation{Institute for Solid State Research, IFW-Dresden, P.O.Box 270116, D-01171 Dresden,
Germany}
\author{A.~Varykhalov}
\author{R.~Follath}
\address{BESSY GmbH, Albert-Einstein-Strasse 15, 12489 Berlin, Germany}
\author{S.~V.~Borisenko}
\affiliation{Institute for Solid State Research, IFW-Dresden, P.O.Box 270116, D-01171 Dresden,
 Germany}

\begin{abstract}\center
\begin{minipage}{\textwidth}\bigskip
\textbf{The distribution of valence electrons in metals usually follows the symmetry of an ionic
lattice. Modulations of this distribution often occur when those electrons are not stable with
respect to a new electronic order, such as spin or charge density waves. Electron density waves
have been observed in many families of superconductors\cite{Wise, Kawasaki, Morosan}, and are often
considered to be essential for superconductivity to exist\cite{Chakravarty}. Recent
measurements\cite{Ding, Kaminski, Shen, Feng, FengSr}  seem to show that the properties of the iron
pnictides\cite{Chen2, Kamihara} are in good agreement with band structure calculations that do not
include additional ordering, implying no relation between density waves and superconductivity in
those materials\cite{Singh0, Ma, Nekrasov, Mazin}. Here we report that the electronic structure of
Ba$_{1-x}$K$_x$Fe$_2$As$_2$ is in sharp disagreement with those band structure
calculations\cite{Singh0, Ma, Nekrasov, Mazin}, instead revealing a reconstruction characterized by
a $(\pi,\pi)$ wave vector. This electronic order coexists with superconductivity and persists up to
room temperature. }
\end{minipage}
\end{abstract}

\maketitle

Calculations of the electronic structure of the new pnictide superconductors unanimously predict a
Fermi surface\,(FS) consisting of hole-like pocket in the centre ($\Gamma$ point) of the Brillouin
zone\,(BZ) and electron-like ones at the corners (X point) of the BZ.  A shift by the $(\pi, \pi)$
vector would result in a significant overlap of these FSs. Such an electronic structure is highly
unstable  since any interaction allowing an electron to gain a $(\pi, \pi)$ momentum would favour a
density wave order, which then results in aforementioned shift and a concomitant opening of the
gaps, thus  strongly reducing the electronic kinetic energy. It is surprising that ARPES data are
reported to be in general, and sometimes in very detailed\cite{Shen}, agreement with the
calculations giving a potentially unstable solution\cite{Kaminski, Ding, FengSr}. Even in the
parent compound, where the spin density wave transition is clearly seen by other
techniques\cite{delaCruz, Huang}, no evidence for the expected energy gap has been detected by
photoemission experiments\cite{Feng, FengSr}. In fact, no consensus exists regarding the overall FS
topology. According to Refs.\,\onlinecite{Ding} and \onlinecite{Kaminski}, there is a single
electron-like FS pocket around the X point, while Ref.~\onlinecite{Zhao} reports two intensity
spots without any discernible signature for the electron pocket in the normal state. Intensity
spots near the X point can also be found in Refs. \onlinecite{Ding}, \onlinecite{FengSr} and
\onlinecite{Shen}, but those are interpreted as parts of electron-like pockets. Obviously, such
substantial variations in the photoemission signal preclude unambiguous assignment of the observed
features to the calculated FS, leaving the electronic structure of the arsenides unclear.

\begin{figure}
\begin{center}
\includegraphics[width=\columnwidth]{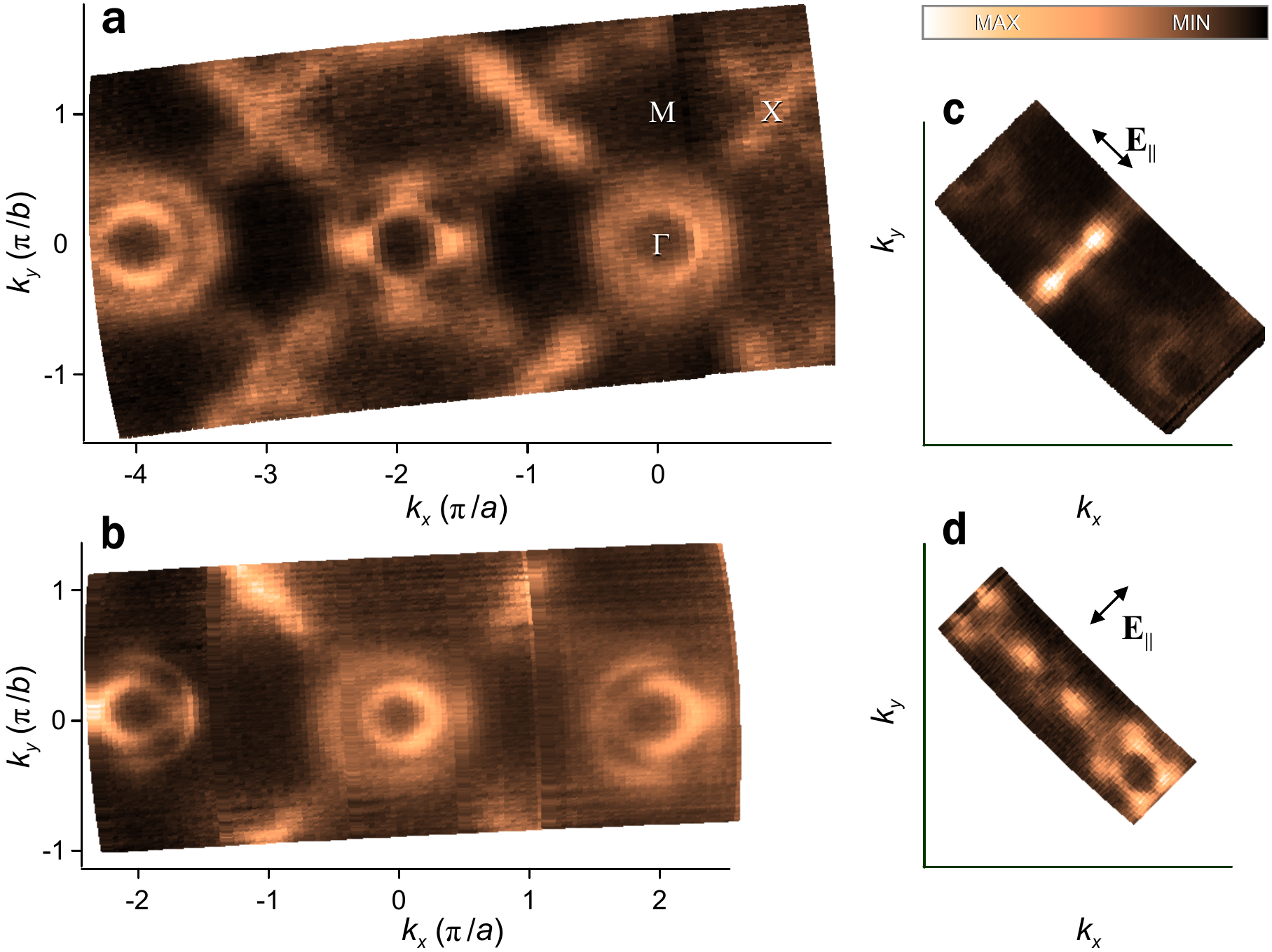}\\
\caption{ Fermi surface topology of Ba$_{1-x}$K$_{x}$Fe$_{2}$As$_{2}$. The colour plots display the
photoelectron intensity distribution  as a function of quasimomentum $k_{||}$ integrated in a small
energy window of 15\,meV around the Fermi level.
 {\textbf{a--b}}, Fermi Surface maps  of Ba$_{1-x}$K$_x$Fe$_2$As$_2$ measured using
excitation energy $h\nu$ = 80 and 50 eV   respectively at  $T$ = 14\,K. Images  {\textbf{c}} and
 {\textbf{d}}, measured with $h\nu$ = 80 eV, demonstrate strong effect of light polarization
on the photoemission from the four small FS's surrounding the X point. The component of electric
vector parallel to the sample surface is shown by a double-headed arrow.  Experimental details
concerning the sample preparation can be found in Supplementary Information.}
\end{center}
\end{figure}

In Fig.\,1 we show experimental FS map of Ba$_{1-x}$K$_x$Fe$_2$As$_2$ (BKFA) measured in
superconducting state. To eliminate possible effects of photoemission matrix elements, as well as
to cut the electronic structure at different $k_z$ values, we have done measurements at several
excitation energies (Fig. 1a--b) and polarizations (Fig. 1c--d). Although there are obvious changes
in the intensities of the features, no signatures indicating $k_z$ dispersion can be concluded.
With this in mind, the apparently different intensity distributions at neighboring $\Gamma$ points
appear unusual. While in the first BZ the two concentric contours are broadly consistent with  band
structure calculations, the ``design wheels'' in the second BZ are at  variance with predicted
hole-like circles. The major discrepancy  with theoretical calculations and ARPES
data\cite{Kaminski, Feng, Ding, Zhao, Shen, FengSr} is observed near the  X point, where, according
to the calculations, one expects a sizeable double-walled electron pocket.  Instead, we observe a
propeller-like structure consisting of five small FS sheets: a pocket, situated directly at X, and
four ``blades'' surrounding it. Panels (c) and (d) show that these FSs are not only well separated
but also have different symmetry.

To examine the topology of the five aforementioned  pockets, we look at momentum distribution of
intensity below and above the Fermi level(FL). As can be seen in Fig.\,2a--c, the size of the
X-centered pocket clearly increases when cutting the electronic structure above the FL, and
decreases for the cut below the FL, which proves its electron-like topology. On the other hand, the
behaviour of the blades surrounding the X point is opposite, which shows that they are hole-like.
In Fig.\,2\,e--i we analyze the band dispersions along the cuts given in Fig.\,2d, once again
supporting these conclusions (see caption and Supplementary Information). Therefore, the observed
topology of the Fermi surface is different from that predicted by band structure calculations.
However, in the following we will argue that the results of the calculations can be reconciled with
the experimental data, provided the system reacts to the predicted nesting instability and an
ordered state develops.

\begin{figure}
\begin{center}
\includegraphics[width=\columnwidth]{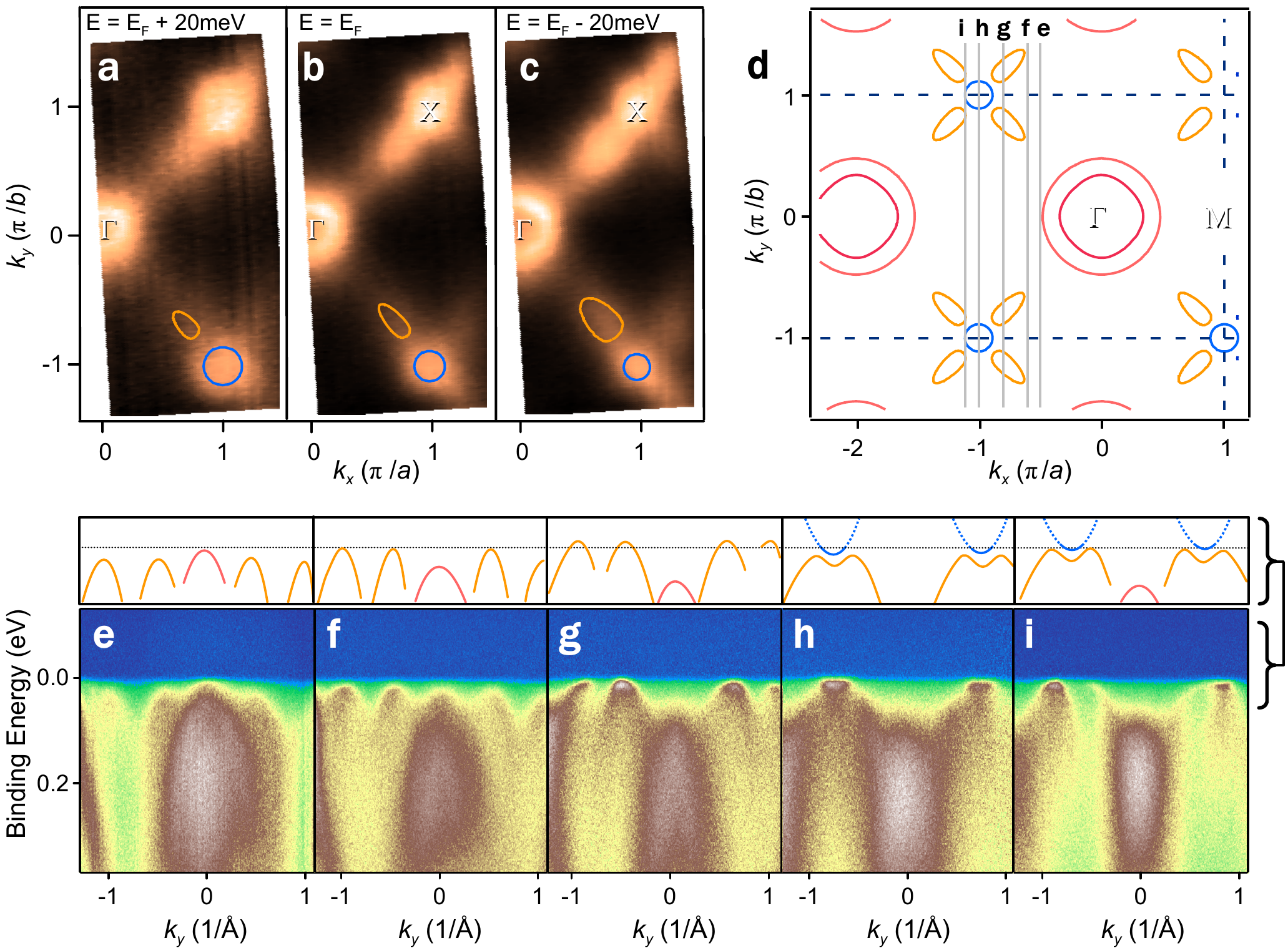}\\
\caption{Low energy electronic structure of Ba$_{1-x}$K$_{x}$Fe$_{2}$As$_{2}$. {\textbf{a--c}},
momentum dependence of the photoemission intensity at constant energy cuts for
Ba$_{1-x}$K$_x$Fe$_2$As$_2$, $T$\,=\,150K, $h\nu = 70$\,eV. The  size of the blade pockets
(outlined by the orange lines) increases with the cut energy $E$ and signals their hole like
topology. The opposite trend for the X-centered pocket (blue circle) implies the electron-like
topology of the pocket. In order not to overlay the experimental data the guide lines are placed
only in the lower part of the images.
 {\textbf{d}}, Summary of the derived FS topology.  {\textbf{e--i}}, Energy-momentum
cuts showing low energy band dispersions ($T\leq 15$K, $h\nu = 80$\,eV), the cartoon on top of each
image schematically outlines the band dispersions and formation of the hole and electron pockets at
the FL. In the cut  {\textbf{e}} one can see five parabolic bands with similar dispersions
approaching, though not crossing, the FL. With movement away from the $\Gamma$ point (cuts
 {\textbf{f--g}}), the four bands on the sides finally cross the FL, forming the blade
structure near the X points. However, when the cut passes directly through the X points (cut
 {\textbf{h}}), these four hole-like bands again dive below the FL, and instead two
electron-like bands, with the curvature clearly opposite to the previous ones, appear from above
the FL.}
\end{center}
\end{figure}

In Fig.\,3 we collate the ARPES data taken at low temperatures for the  parent compound with the
spectra of the doped superconductor, measured using the same light polarisation as in Fig. 1d. The
photoemission intensity distributions shown in panels (a) and (b) are comparable, when taking into
account the difference in the charge carrier concentrations. Moreover, the momentum-energy cuts
(c)--(e) and (f)--(h) show that there is one-to-one correspondence between the underlying band
dispersions in the parent and superconducting samples. Similar locations in the momentum-energy
space, as well as the characteristic, polarisation induced intensity variations clearly suggest
that the blades are also present in BaFe$_2$As$_2$ (BFA). Remarkably, the distance between the
centers of opposite blades tracks the size of the $\Gamma$-FS. Closer inspection of all existing
ARPES data on arsenides confirms an intriguing universality of this observation\cite{Ding, Zhao,
Feng, FengSr, Shen}.

\begin{figure*}
\begin{center}
\includegraphics[width=1.5\columnwidth]{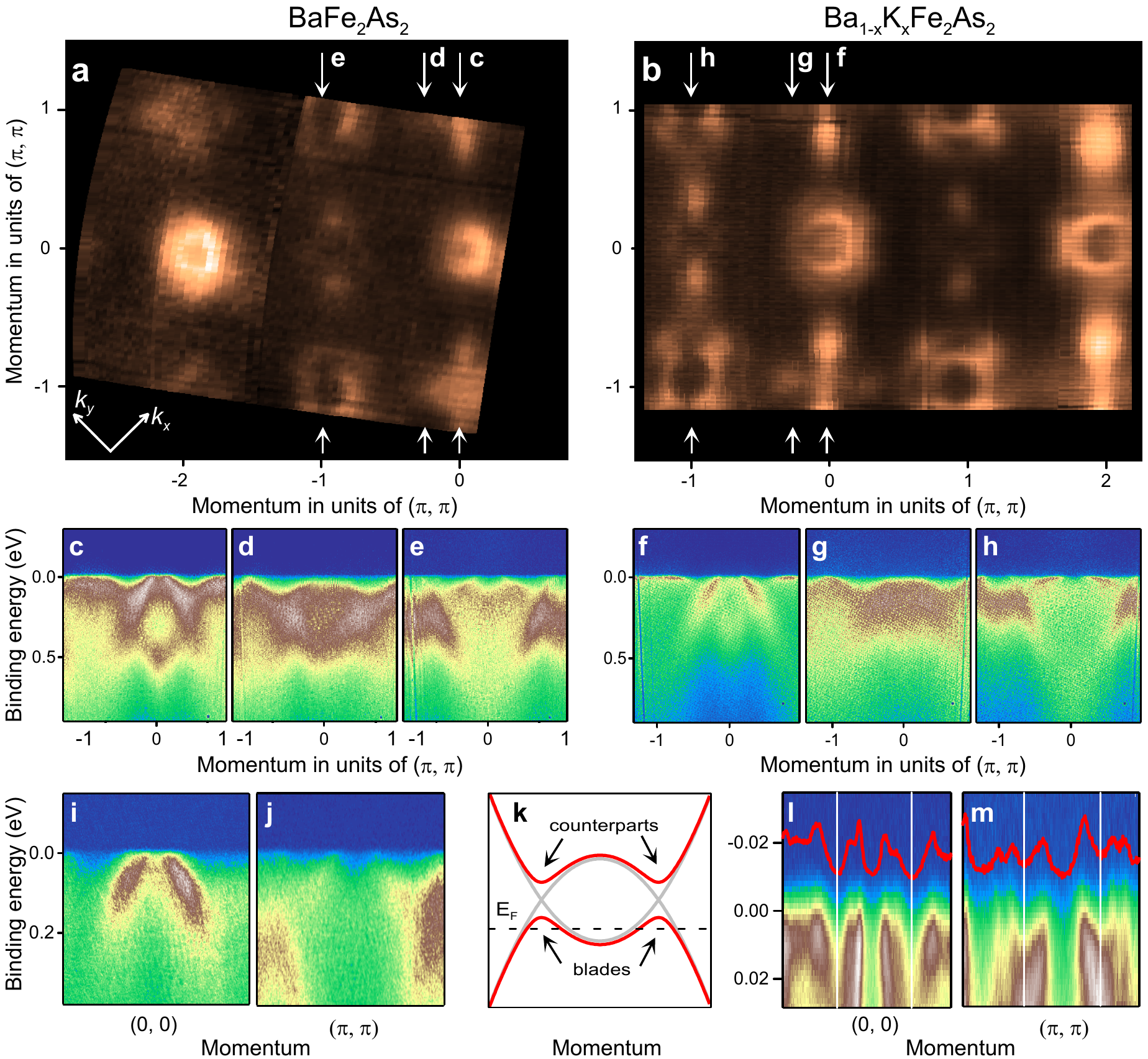}\\
\caption{($\pi$,$\pi$)-reconstruction of the electronic structure.  {\textbf{a--b}}, FS maps for
the BFA and BKFA respectively, $h\nu = 80$\,eV.  {\textbf{c--e}}, {\textbf{f--h}}, Several typical
energy-momentum cuts, normalized to integrated intensity, showing similar band dispersion for BFA
and BKFA. The cuts positions in momentum space are indicated by the arrows in panels
 {\textbf{a}}\,and\, {\textbf{b}}.  {\textbf{i--j}}, Parallel cuts through the
electronic structure of BFA set apart by the $(\pi, \pi)$ vector.  {\textbf{k}}, Simplest model
showing the result of folding of single hole- and electron-like bands.  {\textbf{l--m}}, Parallel
cuts through the $\Gamma$ and X points in electronic structure of BKFA set apart by the $(\pi,
\pi)$ vector. Red curves are the MDCs integrated within the 8 meV showing the symmetric behaviour
about the new BZ boundaries (white lines) due to $(\pi, \pi)$-folding of the original structure.}
\end{center}
\end{figure*}

The origin of the unexpected FS topology near the X point is elucidated by panels (i) and (j) where
momentum-energy cuts separated by the $(\pi, \pi)$ vector are shown: the blades in the parent
compound are created by the interaction of the $(\pi, \pi)$-replica of the $\Gamma$-centered FS
with the X-centered electron-like FS predicted in the calculations. Such a folding implies the
presence of additional ordering that sets in to relax the nesting instability. The formation of the
blades, within this scenario, is illustrated by simple sketch in Fig.\,3k. The intensity of the
$\Gamma$-derived band in panel (j) is lower than that shown in panel (i) due to the weakness of the
scattering potential introduced by the new order as compared to the original crystalline
potential\cite{Brouet, Borisenko}. This is also the case for the FS map in panel (a), where the
ARPES intensities near $\Gamma$ and X points still differ. Similarly, in the superconducting BKFA
this effect prevents immediate detection of the ordered state alluded by the blades. Though,
recovering the band dispersion in the vicinity of the Fermi level as shown in Fig.\,3l--m, the
presence of new BZ boundaries (white lines) can be identified.

An immediate interpretation of the observed similarity in the electronic structures of $x$=0.0 and
$x$=0.3 compounds as shown in Fig.\,3 would be the persistence of  magnetic order in the
superconducting case, which does not contradict  the phase diagram suggested in the Ref.
\onlinecite{Chen}. However, the temperature evolution of the photoemission intensity of the blades
presented in Fig.\,4b rules out any direct connection between the observed FS topology and the
static magnetic order. Despite a noticeably lower spectral weight of the blades as compared to the
$\Gamma$-FS sheets, these hole-like structures  clearly persist to the room temperature, which is
above the structural, magnetic and superconducting transitions in arsenides. Such temperature
behaviour also rules out the possibility that the observed FS topology near the X points is a
consequence of a  pronounced deformation of the calculated band structure, which would destroy the
FS nesting. In order to account for the observed hole-like structures, one would have to shift some
of bands at the X point  by an unlikely 250\,meV\cite{Ma, Mazin, Nekrasov, Kaminski}. In addition,
the potential candidates to form the hole and electron pockets at the X point do not interact due
to symmetry reasons and thus would not be able to account for the picture shown in Fig.\,3k (see
Supplementary Information).

\begin{figure}[b]
\begin{center}
\includegraphics[width=1.0\columnwidth]{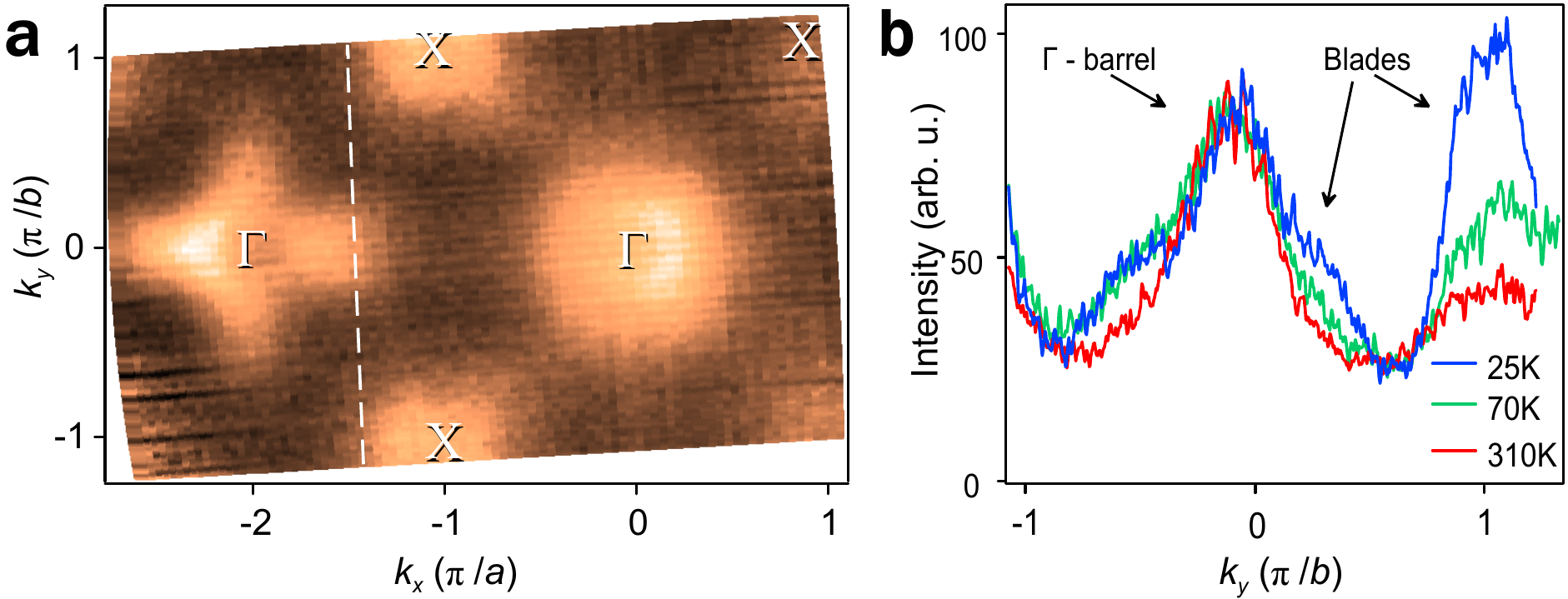}\\
\caption{Blade structure at the X point.  {\textbf{a}}, FS map measured at $T$=300\,K ($h\nu =
50$\,eV) demonstrating persistence of the blades related intensity to high temperatures.
 {\textbf{b}}, Temperature dependence of the blades intensity as compared to the $\Gamma$
barrel. The plot represents the intensity  along the cut schematically shown in panel
 {\textbf{a}} and was integrated in the 100\,meV window centered at the FL.}
\end{center}
\end{figure}

Our ARPES data manifest the presence of  electronic order of a special kind. This order sets in
already at high temperatures and is obviously dictated by the nesting instability predicted in the
calculations. Most likely it is this electronic order which results in the structural transition at
lower temperatures in the parent compound, since in the case of  LaOFFeAs the static magnetic order
develops only after the structural transition and magnetic moments are much smaller than
theoretically expected\cite{delaCruz, Huang}. On the other hand,  presence of the $(\pi, \pi)$
excitations (q=1.15 \AA{$^{-1}$}) above the superconducting transition is indeed implied by the
inelastic neutron scattering data \cite{Christianson} and their transformation into a resonance
below $T_\textup{c}$ could indicate their crucial role in the mechanism of superconductivity. If
these excitations correspond to a fluctuating stripe-like order, the propeller structure can result
from a superposition of  two pairs of blades originating from folding associated with $(\pi, \pi)$
and $(\pi, -\pi)$ wavevectors.

The observed order is not a conventional charge-density wave, since the atoms respond only at
considerably lower temperatures\cite{delaCruz, Huang}, but since the low-lying electronic structure
is formed exclusively by the d-electrons, it may well be related to a more complex order parameter
as in the case of the hidden order scenario suggested for the cuprates \cite{Chakravarty}.
Observation of the reconstruction below $T_\textup{c}$ implies the coexistence of the $(\pi, \pi)$
order and the superconductivity, which is confirmed by opening of the superconducting gaps on the
``propeller'' FS\cite{Danil}.

 The order weakens with doping, which probably explains
that it is no longer strong enough to cause the structural transition and/or static magnetism, but
is sufficient to open smaller $(\pi, \pi)$-gaps near the X point, thus providing a high density of
states at the Fermi level that might be necessary for superconductivity \cite{Radtke}.


\vspace{0.5cm} \noindent{\textbf{Acknowledgements:}}{The project was supported, in part, by the DFG
under Grant No. KN393/4 and BO 1912/2-1. We are grateful to I. Eremin,  O. K. Andersen,  L. Boeri,
I. Mazin and M. Rümelli for fruitful discussions.}

\vspace{0.5cm} \noindent{\textbf{Competing Interests:}} The authors declare that they have no
competing financial interests.

\vspace{0.5cm} \noindent{\textbf{Correspondence:}}Correspondence and requests for materials should
be addressed to S. V. Borisenko (S.Borisenko@ifw-dresden.de).


\begin{thebibliography}{1}

\bibitem{Wise}  Wise, W. D.  \textit{et\,al.}  Charge-density-wave origin of cuprate checkerboard visualized by scanning tunnelling microscopy.
\textit{Nature Physics} \textbf{4}, 696 -- 699 (2008).


\bibitem{Kawasaki}  Kawasaki, S.  \textit{et\,al.}  Enhancing the superconducting transition temperature of CeRh$_{1-x}$Ir$_x$In$_5$
due to the strong-soupling effects of antiferromagnetic spin fluctuations: An 115In nuclear
quadrupole resonance study \textit{Phys. Rev. Lett.} \textbf{96}, 147001 (2006).


\bibitem{Morosan}  Morosan, E. \emph{et al}. Superconductivity in Cu$_x$TiSe$_2$ \textit{Nature Physics}
\textbf{2}, 554 -- 550 (2006).


\bibitem{Chakravarty} Chakravarty, S. \textit{et al.} Hidden order in the cuprates. \textit{Phys. Rev. B} \textbf{63}, 094503 (2001).

\bibitem{Kaminski}          Liu, C. \textit{et\,al.}  The Fermi surface of Ba$_{1-x}$K$_{x}$Fe$_2$As$_2$ and its evolution with doping.
\textit{arXiv:}0806.3453v1.

\bibitem{Ding}              Ding, H.  \textit{et\,al.}  Observation of Fermi-surface-dependent nodless superconducting gaps in Ba$_{0.6}$K$_{0.4}$Fe$_2$As$_2$. \textit{Europhys. Lett.} \textbf{83}, 47001
(2008).


\bibitem{FengSr}            Zhang, Y.  \textit{et\,al.}
 Correlation effects of exchange splitting and coexistence of spin-density-wave
 and superconductivity in single crystalline Sr$_{1-x}$K$_x$Fe$_2$As$_2$.
\textit{arXiv:}0808.2738v1.

\bibitem{Feng}              Yang, L. X.  \textit{et\,al.}  Electronic structure and exotic exchange splitting in spin-density-wave states of BaFe$_2$As$_2$.
\textit{arXiv:}0806.2627v2.


\bibitem{Shen}          Lu, D. H. \textit{et\,al.} Electronic structure of the iron-based superconductor
LaOFeP. \textit{Nature\,(London)} \textbf{455}, 81 -- 84 (2008).


\bibitem{Chen2} Chen, X. H.,  Wu, T.,  Wu, G.,   Liu,  R. H.  \&  Fang,  D. F.  Superconductivity at 43 K in SmFeAsO$_{1-x}$F$_x$. \textit{Nature\,(London)} \textbf{453}, 761 -- 762
(2008).


\bibitem{Kamihara}   Kamihara, Y.,  Watanabe, T.,  Hirano, M.,  \& Hosono, H. Iron-based layered
superconductor La[O$_{1-x}$F$_x$]FeAs ($x$ = 0.05 -- 0.12) with T$_\textup{c}$ = 26 K.  \textit{J.
Am. Chem. Soc.} \textbf{130}, 3296 -- 3297 (2008).



\bibitem{Singh0}           Singh, D. J.  Electronic structure and doping in BaFe$_2$As$_2$ and LiFeAs:  density functional
calculations. \textit{ arXiv:}0807.2643v1.


\bibitem{Ma}             Ma, F.,  Lu, Zh.-Y. \&  Xiang, T.   Electronic band structure of BaFe$_2$As$_2$.
\textit{arXiv:}0806.3526v1.


\bibitem{Nekrasov}       Nekrasov, I.A., Pchelkina, Z.V. \&  Sadovskii, M.V. Electronic structure of prototype AFe$_2$As$_2$ and ReOFeAs high-temperature superconductors: a
comparison. \textit{arXiv:}0806.2630v1.

\bibitem{Mazin}         Mazin,  I. I., Singh, D. J., Johannes, M. D. \&  Du, M.H. Unconventional sign-reversing superconductivity in LaFeAsO$_{1-x}$F$_x$. \textit{Phys. Rev. Lett.} \textbf{101}, 057003
(2008).


\bibitem{delaCruz}      de la Cruz, C. \textit{et\,al.} Magnetic order close to superconductivity in the
iron-based layered LaO$_{1-x}$F$_x$FeAs systems. \textit{Nature\,(London)} \textbf{453}, 899 -- 902
(2008).


\bibitem{Huang}         Huang, Q. \textit{et\,al.}  Magnetic order in BaFe$_2$As$_2$, the parent compound of the FeAs based
superconductors in a new structural family. \textit{arXiv:}0806.2776v2.


\bibitem{Zhao}              Zhao,  L. \textit{et\,al.}  Unusual superconducting gap in (Ba,K)Fe$_2$As$_2$.
\textit{arXiv:}0807.0398v1.

\bibitem{Brouet}    Brouet, V. \textit{et\,al.} Angle-resolved photoemission study of the evolution of band structure and charge density wave
properties in $R$Te$_3$ ($R$=Y, La, Ce, Sm, Gd, Tb, and Dy).  \textit{Phys. Rev. Lett.}
\textbf{77}, 235104 (2008)

\bibitem{Borisenko}      Borisenko, S. V. \textit{et\,al.}  Pseudogap and charge density waves in two dimensions. \textit{Phys. Rev. Lett.} \textbf{100}, 196402
(2008).


\bibitem{Chen}          Chen,  H. \textit{et\,al.}  Coexistence of the spin-density-wave and superconductiviy in the Ba$_{1-x}$K$_x$Fe$_2$As$_2$.
\textit{arXiv:}0807.3950v1.


\bibitem{Christianson}  Christianson, A. D.  \textit{et\,al.}   Resonant Spin Excitation in the High Temperature Superconductor
Ba$_{0.6}$K$_{0.4}$Fe$_2$As$_2$. \textit{arXiv:}0807.3932v1.


\bibitem{Danil} Evtushinky, D. V. \textit{et al.} Momentum dependence of the superconducting gap in Ba$_{1-x}$K$_{x}$Fe$_2$As$_2$. \textit{arXiv:}0809.4455v1.

\bibitem{Radtke}        Radtke, R. J.  and Norman, M. R. Relation of extended Van Hove singularities to high-temperature superconductivity
within strong-coupling theory. \textit{Phys. Rev. B} \textbf{50}, 9554 -- 9560 (1994).



\end{thebibliography}
\end{document}